\documentclass[prd,nofootinbib,superscriptaddress,showpacs,showkeys]{revtex4}
\usepackage[a4paper, hdivide={2.5cm,,2.5cm}, vdivide={3cm,,3cm}]{geometry}
\usepackage{amstext}
\usepackage[english]{babel}
\usepackage[intlimits]{amsmath}
\usepackage{amssymb}
\usepackage{amsthm}
\usepackage{graphicx}
\usepackage[ansinew]{inputenc}
\usepackage{slashed} 
\usepackage{units}
\usepackage{hyperref}
\hypersetup{
    pdftitle={Neutrino Hierarchies from a Gauge Symmetry},    
    pdfauthor={Julian Heeck, Werner Rodejohann}     
}



\begin{document}

\title{Neutrino Hierarchies from a Gauge Symmetry}

\author{Julian \surname{Heeck}}
\email{julian.heeck@mpi-hd.mpg.de}
\affiliation{Max--Planck--Institut f\"ur Kernphysik, Saupfercheckweg 1, 69117 Heidelberg, Germany}
\affiliation{Institute for Theoretical Physics, Kanazawa University, Kanazawa 920-1192, Japan}

\author{Werner \surname{Rodejohann}}
\email{werner.rodejohann@mpi-hd.mpg.de}
\affiliation{Max--Planck--Institut f\"ur Kernphysik, Saupfercheckweg 1, 69117 Heidelberg, Germany}

\begin{abstract}
We consider the phenomenology of the gauged abelian symmetry $B+ 3 (L_e - L_\mu - L_\tau)$. Right-handed neutrinos necessary to cancel triangle anomalies are used in a type-I seesaw scheme to create active neutrino masses. Breaking the $B+ 3 (L_e - L_\mu - L_\tau)$ symmetry spontaneously below the seesaw scale generates low energy neutrino mass matrices with the approximate symmetries $L_e$ (leading to normal hierarchy) or $L_e - L_\mu - L_\tau$ (inverted hierarchy). For the latter we need to introduce a $\mathbb{Z}_2$ symmetry which decouples one of the right-handed neutrinos. If exact, this $\mathbb{Z}_2$ leads to a Majorana dark matter candidate that interacts with the Standard Model via the $Z'$ and a scalar $s$ originating from spontaneous breaking of the new symmetry. The measured relic abundance of the dark matter particle can be obtained around the scalar and $Z'$ resonances, while direct detection experiments are mainly sensitive to scalar exchange, which is induced by mass mixing of $s$ with the standard Higgs.
\end{abstract}

\maketitle

\newcommand{\dd}{\mathrm{d}} 
\newcommand{\del}{\partial}
\let \arrowvec \vec
\def \vec#1{{\boldsymbol{#1}}}
\let \Lold \L
\def \L {\mathcal{L}} 
\newcommand{\matrixx}[1]{\begin{pmatrix} #1 \end{pmatrix}} 
\newcommand{\tr}{\mathrm{tr}}
\newcommand{\sm}{\mathrm{SM}}
\newcommand{\diag}{\mathrm{diag}}
\newcommand{\hc}{\mathrm{h.c.}}
\newcommand{\Le}{L_e}
\newcommand{\Lemt}{\overline{L}}
\newcommand{\Lmt}{L_\mu-L_\tau}
\newcommand{\ra}{\rightarrow}
\def \Z2{\mathbb{Z}_2}

\let \oldepsilon \epsilon
\def \epsilon{\varepsilon}


\section{Introduction}
The observed neutrino mixing angles and mass-squared differences have
launched an avalanche of models trying to explain their values. One
possible starting point is the Majorana mass matrix in flavor basis, on
which one then imposes symmetries. As far as continuous abelian
symmetries of lepton numbers go, three interesting linear combinations
have been identified for the zeroth-order approximation: $\Le$, $\Lemt
\equiv L_e - L_\mu - L_\tau$ and $\Lmt$, leading to normal hierarchy
(NH), inverted hierarchy (IH) and quasi-degeneracy in the neutrino
mass spectrum~\cite{Choubey:2004hn}. The flavor structure of the mass
matrices is 
\begin{align}
	\mathcal{M}_\nu^{\Le} \sim \matrixx{0 & 0 & 0\\ 0 & \times & \times \\ 0 & \times & \times} , &&
	\mathcal{M}_\nu^{\Lemt} \sim \matrixx{0 & \times & \times \\ \times & 0 & 0\\ \times & 0 & 0} , &&	
	\mathcal{M}_\nu^{\Lmt} \sim \matrixx{\times & 0 & 0 \\ 0 & 0 &
\times \\ 0 & \times & 0} ,
	\label{eq:symmetricforms}
\end{align}
where $\times$ denotes a non-zero entry. 
The last matrix conserving $\Lmt$ has the interesting
property of being anomaly-free~\cite{zero} (in the Standard Model (SM) with
massless neutrinos), so the symmetry can be gauged, which leads to
numerous interesting effects~\cite{mu-tau-short}. The other two
symmetries have been considered as global
symmetries~\cite{Lemt,lowLe,Lemt2} (or as an anomalous $U(1)$~\cite{anomalousU1}), but no effort has been put forward
to construct a local version. 

The reason why it is not easily possible to promote $\Le$ or $\Lemt$ to a
local symmetry with the SM particle content is due to arising quantum
anomalies, even if we introduce SM-singlet right-handed neutrinos
(RHNs). Extending the chiral fermion content of the model could serve as a
viable way to cancel these triangle anomalies and construct a
renormalizable Lagrangian. In the case of the above symmetries it can
be shown that a complete fourth generation of fermions suffices to
accomplish this task.\footnote{This is easy to see because $L_e-L_\mu-L_\tau+L_{\ell_4}$ is a vectorlike symmetry in analogy to $L_\alpha-L_\beta$.} However, the strict bounds on the
fourth-generation fermions complicate model building severely,
especially when it comes to the mass matrix of the---then
four---active neutrinos. 

A different way to cancel anomalies is the modification of the
symmetry itself. For example, the quantum number $B- 3 L_e$ is anomaly
free in the SM plus RHNs~\cite{B-3Le}, and leads to an $\Le$ symmetric
neutrino mass matrix for the right-handed neutrinos (see Eq.~\eqref{eq:symmetricforms}). Models based on symmetries of the type
$B - \sum_\ell x_\ell L_\ell$ and $\sum_\ell y_\ell L_\ell$ (with the
constraints $\sum_\ell x_\ell = 3$ and $\sum_\ell y_\ell = 0$, respectively) have
been discussed for example in
Refs.~\cite{B-stuff,Chang:2000xy,Davoudiasl:2011sz,Salvioni:2009jp}. 
Seesaw neutrino models with an additional $U(1)'$ are also discussed
in Ref.~\cite{Chen:2011de}.  
Some of the phenomenological aspects of such models (the scalar
sector, dark matter candidates, etc.) are similar to 
frequently discussed $B-L$ analyses. However, choosing gauge groups
that include flavor information makes it possible to provide
predictions on neutrino mixing and mass spectrum, which is impossible 
in theories based solely on $B-L$. This interesting connection of
flavor and gauge physics motivates us here to discuss the
minimal gauged $B+3\Lemt$ model, which is free of anomalies if
right-handed neutrinos with proper charges under the new $U(1)'$ are
introduced. Active neutrino masses are a result of a type-I seesaw
mechanism, which is applicable only in the case of a broken symmetry,
because the zeroth order right-handed neutrino mass matrix obeying
$\Lemt$ symmetry has rank 2. 
Interestingly, the resulting low energy neutrino mass
matrix $\mathcal{M}_\nu$ does not necessarily obey a flavor structure 
resembling the one required from $\Lemt$ conservation. Indeed, in what
follows we will see that details of the breaking of 
$B+ 3 (L_e - L_\mu - L_\tau)$ can lead to low energy neutrino
physics with a normal or inverted hierarchy for the active
neutrinos. As a possibility to force the inverted hierarchy in the
active neutrino sector, we introduce a $\mathbb{Z}_2$
symmetry. Interestingly, the very same $\mathbb{Z}_2$ turns out to
render one of the right-handed neutrino stable, and to become a dark
matter candidate.

The paper is build up as follows: we will show the anomaly freedom of
our symmetry in Sec.~\ref{sec:anomalies}. In
Sec.~\ref{sec:neutrino_masses} we show that the model can lead to either
$\Le$ or $\Lemt$ symmetric low-energy neutrino mass matrices via
type-I seesaw, depending on the number of right-handed neutrinos and
additional discrete symmetries.
The $Z'$ phenomenology of $B+ 3 \Lemt$ has already been briefly
considered in Ref.~\cite{Lee:2010hf}, as a special case of $B -
\sum_\ell x_\ell L_\ell$, so we devote only a small section to its
discussion (Sec.~\ref{sec:gauge_sector}). Since the minimal scalar
sector consists only of one additional complex scalar, the effects are
well known from, e.g., minimal $B-L$ models. The resulting mixing
among the scalars is briefly derived in
Sec.~\ref{sec:scalar_sector}. The discussion of the neutrino mass
matrix naturally leads to an additional exchange symmetry,
which---properly implemented---yields a stable right-handed neutrino
as a dark matter candidate. We discuss the relic abundance of said
dark matter candidate around the scalar and $Z'$ resonances in
Sec.~\ref{sec:dark_matter}. We conclude our findings in
Sec.~\ref{sec:conclusion}. The calculation of anomalies is presented
in App.~\ref{app:anomalies}, while appendices~\ref{app:Z2breaking}
and~\ref{app:majorana} contain brief calculations that are not of
utter importance to follow the main text.

\section{Gauged \texorpdfstring{$\boldsymbol{ B + 3 (L_e-L_\mu-L_\tau) }$}{B + 3 (Le-Lmu-Ltau)} Symmetry}
\label{sec:anomalies}

We introduce $n_N$ right-handed neutrinos $N_i$ with $U(1)_{B+ 3
\Lemt}$ quantum numbers $Y' (N_i)$. The gauge group representations of
the first-generation fermions are shown in
Tab.~\ref{tab:quantum_numbers}, for the second and third generation
the $U(1)_{B+ 3 \Lemt}$ charge of the leptons changes sign. 
Defining for simplicity $U(1)' \equiv U(1)_{B+3\Lemt}$ and $Y' \equiv
B + 3 \Lemt$ we can calculate the triangle anomalies of the model. As
shown in App.~\ref{app:anomalies}, the model is free of anomalies as
long as the quantum numbers of the right-handed neutrinos satisfy 
\begin{align}
	\sum_{i}^{n_N} Y' (N_i) = -3 \,, &&
	\sum_{i}^{n_N} {Y'}^3 (N_i) = -3^3\,.
\label{eq:righthandedanomalies}
\end{align}
The minimal anomaly free model consists of only one right-handed
neutrino with $\Lemt$ charge $-3$. There are no real solutions of
Eq.~\eqref{eq:righthandedanomalies} for $n_N = 2$, but for odd $n_N$
we can choose $\Lemt (N_{R, 1})= -3$ and add pairs of right-handed
neutrinos with arbitrary---but opposite---charge. Solutions without a
charge $\pm 3$, and therefore without Dirac coupling to the active
neutrinos, can be obtained with five right-handed neutrinos, e.g.~with
the $\Lemt$ charges $-1$, $2$, $-5$, $-5$ and $6$, respectively. It is
clear that in this case $m_D = 0 = \mathcal{M}_R$, and hence the massless
right-handed neutrinos decouple unless $\Lemt$ is broken in a very specific
way. Since this is cumbersome, we will restrict ourselves to RHNs with
charges $\pm 3$ in the following. 

\begin{table}[bt]
\centering
\begin{tabular}[t]{|l|l|l|}
\hline                               
     $L_e = \matrixx{\nu\\e}_L \sim (\vec{1},\vec{2},-1)(3)$ & $e_R^c \sim (\vec{1},\vec{1},+2)(-3)$ & $N_i^c \sim (\vec{1}, \vec{1},0)(Y' (N_i^c))$\\
     \hline
     $Q^u_L = \matrixx{u\\ d}_L \sim (\vec{3},\vec{2},+\frac{1}{3})(+\frac{1}{3})$ & $u_R^c \sim (\vec{\overline{3}},\vec{1},-\frac{4}{3})(-\frac{1}{3})$ & $d_R^c \sim (\vec{\overline{3}},\vec{1},+\frac{2}{3})(-\frac{1}{3})$ \\
\hline 
\end{tabular}
\caption{$SU(3)_C\times SU(2)_L\times U(1)_Y \times U(1)_{B+ 3( L_e -
L_\mu - L_\tau)}$ representations of left-handed SM fermions (only
first generation shown) and right-handed neutrinos $N_i$. For the
second and third generation the $U(1)_{B+ 3 \Lemt}$ charge of the
leptons changes sign.} 
\label{tab:quantum_numbers}
\end{table}

The symmetry $U(1)_{B+3\Lemt}$ was already discussed in
Ref.~\cite{Lee:2010hf}, where it is proposed as an origin for R-parity
that also forbids proton decay via higher dimensional operators like
$QQQ L$, which conserve $B-L$ but violate $B-\sum_\ell x_\ell L_\ell$
if $x_\ell \neq 1$. Since in Ref.~\cite{Lee:2010hf} it is only briefly
mentioned that the mass matrix for the right-handed neutrinos has no
vanishing entries in the broken case, we feel it is still worth
discussing the neutrino masses in more detail, due to their interesting structure. 

It should be stressed that even though we are taking a non-supersymmetric model for simplicity, a similar discussion holds for the supersymmetric case of Ref.~\cite{Lee:2010hf}. Supersymmetric particles aside, the main difference is the need for a second complex scalar (super-)field to fill the vanishing entries in the neutrino mass matrix. The model (superpotential, mass spectrum etc.) is then similar to supersymmetric $B-L$ models, which are intensively discussed in e.g.~Refs.~\cite{moreSUSYB-L}. Assuming similar vacuum expectation values for both scalars makes the discussion of neutrino masses identical to Sec.~\ref{sec:neutrino_masses}. The scalar and dark matter sectors will of course differ from Sec.~\ref{sec:scalar_sector} and Sec.~\ref{sec:dark_matter} in a supersymmetric context. For example, the mixing of the Higgs doublets $H_i$ and the new scalars $S_i$ will be severely suppressed~\cite{SUSYB-L}, making the $Z'$ boson the main mediator between the dark matter and SM sector. A discussion of dark matter (especially in the context of the additional $\Z2$ symmetry that leads to inverted hierarchy) would be interesting, but lies outside the realm of this work.

\section{Neutrino masses}
\label{sec:neutrino_masses}

In this section we discuss various interesting possibilities of the
new gauge symmetry in the neutrino sector.

\subsection{Three right-handed neutrinos}

The most natural quantum number assignment for three right-handed neutrinos that cancels the anomalies of Eq.~\eqref{eq:righthandedanomalies} is $+3$, $-3$ and $-3$. After electroweak symmetry breaking, the Dirac and (symmetric) Majorana mass matrices for $\overline{\nu}_i N_j$ and $\overline{N}^c_i N_j$, respectively, take the form
\begin{align}
	m_D = \matrixx{a & 0 & 0\\ 0 & b & c\\ 0 & d & e} , &&
	\mathcal{M}_R = \matrixx{0 & X & Y\\ \cdot & 0 & 0 \\ \cdot & \cdot & 0} .
	\label{eq:mass_matrices}
\end{align}
As already mentioned in the introduction, the matrix $\mathcal{M}_R$ is singular, which means the usual seesaw formula $\mathcal{M}_\nu \simeq - m_D \mathcal{M}_R^{-1} m_D^T$ for the light neutrinos in the limit $X,Y \gg (m_D)_{ij}$ is not applicable. Instead of the $3\,\nu_\mathrm{light} + 3\, \nu_\mathrm{heavy}$ scheme known from seesaw, the diagonalization of the full $6\times 6$ matrix leads to the hierarchy $2\,\nu_\mathrm{heavy} + 2 \,\nu_\mathrm{electroweak} + 2\, \nu_\mathrm{light}$, which is clearly not in agreement with experiments.

Since the model looks quite different after $U(1)'$ breaking, let us introduce a complex scalar field $S\sim (\vec{1},\vec{1},0)(+6)$ which acquires a vacuum expectation value (VEV). Collider limits on additional heavy neutral gauge bosons typically give limits $M_{Z'} / g' \sim \langle S \rangle \gtrsim 1$--$\unit[10]{TeV}$ for the VEV, to be further discussed in Sec.~\ref{sec:gauge_sector}. $S$~couples to the right-handed neutrinos, so $\langle S \rangle$ fills all texture zeros in $\mathcal{M}_R$. As a result, $\mathcal{M}_R$ is in general an invertible matrix after $B+3\Lemt$ breaking:
\begin{align}
	\mathcal{M}_R = \matrixx{A & X & Y\\ \cdot & B & C \\ \cdot & \cdot & D}, &&
	\mathcal{M}_R^{-1} = -\frac{1}{\det \mathcal{M}_R}\, \matrixx{C^2 - B D & D X - C Y & B Y - C X\\ \cdot & Y^2 - A D & A C - X Y \\ \cdot & \cdot & X^2 - A B} .
\end{align}
The scaling $X,Y \gg \langle S \rangle \gg (m_D)_{ij}$ leads to the order-of-magnitude structure of the low-energy neutrino mass matrix
\begin{align}
	\mathcal{M}_\nu \simeq - m_D \mathcal{M}_R^{-1} m_D^T \sim \matrixx{\epsilon^2 & \epsilon & \epsilon\\ \cdot & 1 & 1\\ \cdot & \cdot & 1},
	\label{eq:seesaw_oom}
\end{align}
with $\epsilon \equiv \langle S \rangle / X$. Consequently, a low $B+
3 \Lemt$ breaking scale (compared to $\mathcal{M}_R$, not $m_D$)
$\epsilon \sim 0.1$ actually leads to a mass matrix that approximately
conserves $\Le$ instead of $\Lemt$~(see Eq.~\eqref{eq:symmetricforms}). 
It is however not the most general $\Le$ symmetric matrix, because the
zeroth-order mass matrix has the structure 
\begin{align}
 \mathcal{M}_\nu \sim \matrixx{0 & 0 & 0\\ \cdot & (c X - b Y)^2 & (c X - b Y) (e X - d Y)\\ \cdot & \cdot & (e X - d Y)^2} + \mathcal{O}(\epsilon)\,,
\end{align}
which gives only one massive neutrino $\nu_3 \sim  (c X  - b Y)\,
\nu_\mu + (e X - d Y)\, \nu_\tau$.

The solar mixing angle is still undefined at this
order, due to an accidental $O(2)$ symmetry of the matrix (see
Ref.~\cite{hiddenO2}). Since the symmetry allows for mixing of $\mu$
and $\tau$, the charged lepton mass matrix is not diagonal in general
and contributes to $\theta_{23}$. The atmospheric mixing angle will
therefore be a combination of the charged-lepton mixing and the
neutrino one 
\begin{align}
\tan \theta_{23}^\nu \simeq \frac{c X  - b Y}{e X - d Y}\,,
\end{align}
so we expect large but non-maximal mixing.

\begin{figure}[t]
	\begin{center}
		\includegraphics[width=0.48\textwidth]{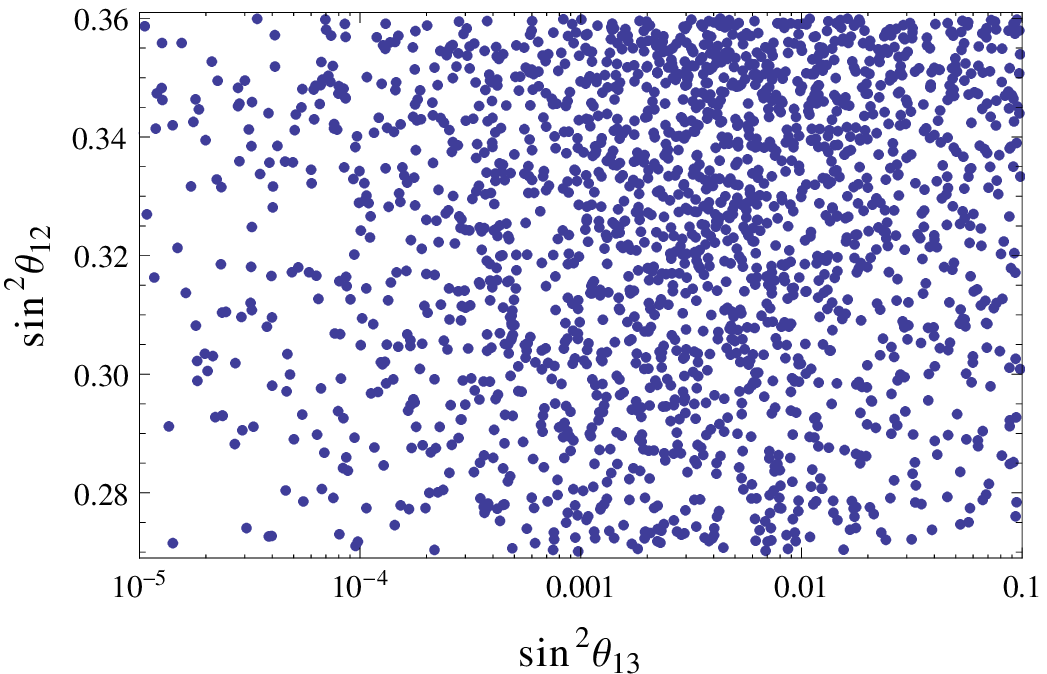}\hspace{1ex}
		\includegraphics[width=0.48\textwidth]{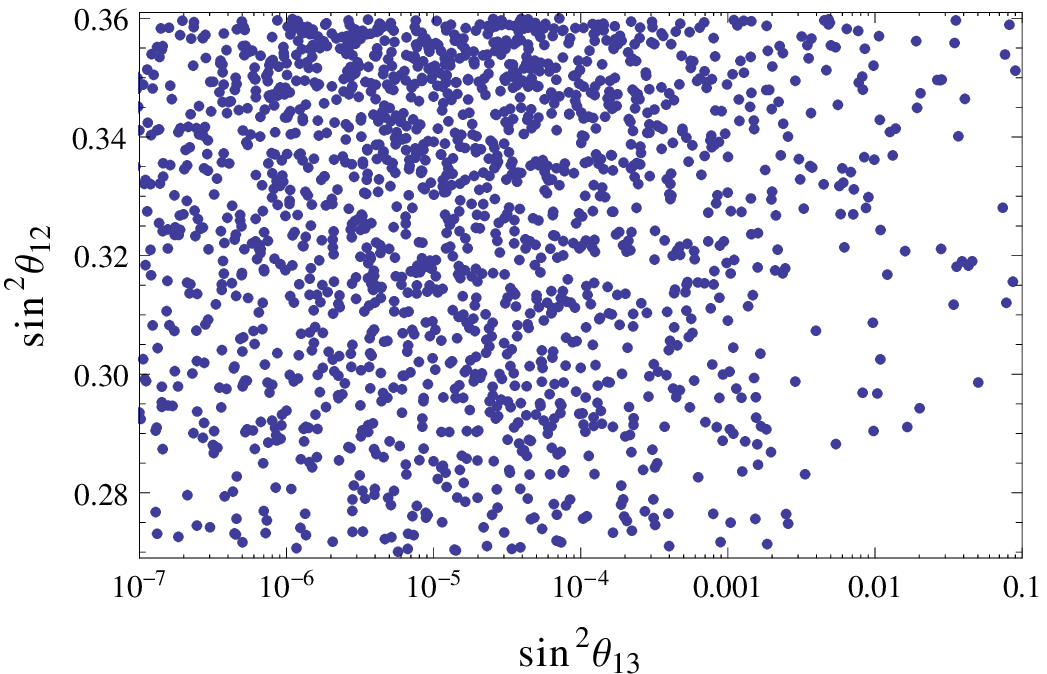}
	\end{center}
		\caption{Left: scatter plots using the broken $B+3 \Lemt$ low-energy neutrino mass matrix~\eqref{eq:seesaw_oom} ($\epsilon = 0.05$) that leads to NH. Right: scatter plots using the neutrino mass matrix~\eqref{eq:fiveplusZ2} ($\epsilon = 0.1$) with five right-handed neutrinos and a $\mathbb{Z}_2$, which leads to IH. The accepted values of the mixing parameters satisfy the $3\sigma$ bounds from Ref.~\cite{theta13}, except for $\theta_{23}$, because it can be arbitrarily adjusted by the charged-lepton contribution.}
	\label{fig:scatter}
\end{figure}

While not particularly predictive, we show the distribution of the
mixing angles $\theta_{12}$ and $\theta_{13}$ in
Fig.~\ref{fig:scatter} (left). For these we generated random Yukawa
couplings $|(m_D)_{ij}|\leq 1$, $|A|,|B|,\dots <\epsilon$ and
$|X|,|Y|>1$ that lead to neutrino mixing parameters in their $3\sigma$
range~\cite{theta13}. Here and in the following we restrict the
parameters to real values for simplicity, resulting in vanishing
CP-violating phases in the mixing matrix. In any case, since the Yukawa couplings can have arbitrary phases, we do not expect our model to be able to predict the CP-violating phases. The solar angle tends to be
large while the reactor angle $\theta_{13}$ is generally small, but in good
agreement with the recent T2K~\cite{t2k},
Double-Chooz~\cite{doublechooz}, Daya Bay~\cite{dayabay} and RENO~\cite{reno} results of $\sin^2 \theta_{13} \simeq 0.025$--$0.03$.

The units of $m_D$ and $\mathcal{M}_R$ have not been specified yet, because they only fix the overall neutrino mass scale---and hence the $\Delta m_{ij}^2$---but not the mixing angles. In the usual seesaw manner, the magnitude $m_D^2/\mathcal{M}_R \simeq \unit[0.1]{eV}$ does not fix the seesaw scale, but naturalness hints at a high scale.

Since the NH structure~\eqref{eq:seesaw_oom} has already been
recognized before as a byproduct of the softly broken global $\Lemt$
(see for example Refs.~\cite{lowLe}), we will not attempt to redo all
the calculations done before. Instead, we construct a model with a
true $\Lemt$ symmetry (and therefore inverted hierarchy) instead of
the effective $L_e$ as above. It turns out we are just a $\Z2$
symmetry away. 

\subsection{Three right-handed neutrinos and \texorpdfstring{$\Z2 $}{Z2} symmetry}
\label{sec:threeplusZ2}

The reason for the different approximate symmetries in $\mathcal{M}_R$
and $\mathcal{M}_R^{-1}$ is the occurring vanishing eigenvalue of
$\mathcal{M}_R$ in the unbroken case. Since
the number of right-handed neutrinos is fixed by anomaly cancellations
to be odd, we cannot simply remove one of the $N_i$ to make
$\mathcal{M}_R$ invertible. We can however forbid its coupling to all
other particles by means of an additional discrete symmetry. We will
discuss the simplest example below. 

Defining an additional $\Z2$ symmetry under which $N_3$ transforms as
$N_3 \ra - N_3$ while all other fields are even,\footnote{This is
equivalent to an exchange symmetry $N_2\leftrightarrow N_3$ as can be
seen using the basis $\Psi_1 \sim N_2 + N_3$, $\Psi_2 \sim N_2 -
N_3$.} the only allowed interactions for $N_3$ are 
\begin{align}
\begin{split}
	\L_{N_3} &= i \overline{N_3} \gamma^\mu \left( \del_\mu - i (-3)  g' Z'_\mu\right) N_3 - Y_\chi S\, \overline{N_3}^c N_3 +\hc \\
	&= \frac{i}{2} \chi^T \mathcal{C} \gamma^\mu  \del_\mu \chi
-\frac{3}{2} g' Z'_\mu \chi^T \mathcal{C}  \gamma^\mu \gamma_5 \chi 
- Y_\chi \frac{v_S}{\sqrt{2}}\ \chi^T \mathcal{C} \chi \left( 1+ \frac{s}{v_S}\right)\,,
\end{split}
	\label{eq:N3lagrangian}
\end{align}
making it stable and heavy after $B+3\Lemt$ breaking. In the last line
we replaced the right-handed Dirac fermion $N_3$ by a Majorana fermion
$\chi$ (see App.~\ref{app:majorana}) and used unitary gauge to make
the $Z'$ boson massive and eliminate $\mathfrak{Im}(S)$. The stable
Majorana fermion $\chi$~is therefore a candidate for dark matter, to
be further examined in Sec.~\ref{sec:dark_matter}. Note that the
stability is due to the $\Z2$, which was introduced to implement an
inverted hierarchy for the active neutrinos. 

The active neutrinos then couple only to $N_1$ and $N_2$, so at most
two active neutrinos acquire mass at tree level. The $B+3\Lemt$
symmetry is broken in $\mathcal{M}_R$ by the parameters $A$ and $B$,
so with the usual seesaw mechanism we find 
\begin{align}
	\mathcal{M}_\nu \simeq - \matrixx{a & 0\\ 0 & b\\ 0 & c} \matrixx{A & X\\ X & B}^{-1} \matrixx{a & 0 & 0\\ 0 & b & c} = \frac{1}{X^2 - A B} \matrixx{ a^2 B & - a b X & - a c X\\ \cdot & b^2 A & b c A\\ \cdot &\cdot & c^2 A},
	\label{eq:Lemtsymmetric}
\end{align}
which features an interesting structure~\cite{Lemt2,Goswami:2008rt}:
The decoupling of $N_3$ results in an invertible
$\mathcal{M}_R^{2\times 2}$, so $\mathcal{M}_\nu$ conserves $L_e -
L_\mu - L_\tau$ in the limit $A,B\rightarrow 0$. This model also gives
a simple explicit realization of ``scaling''~\cite{scaling}, seeing as
the second and third column of $\mathcal{M}_\nu$ are
proportional. Therefore we have an inverted hierarchy solution with
$\theta_{13}=0$, whereas the atmospheric mixing angle is once again
large but random, also due to the contributions of the charged
leptons. At $2$-loop level radiative corrections induce a non-zero
$\theta_{13}$, but of practically 
irrelevant magnitude~\cite{Ray:2010fa}. The solar mixing angle becomes
maximal for $A,B\rightarrow 0$, so the breaking scale needs to be
close to the bare mass terms to lower $\theta_{12}$. 

Since a vanishing reactor angle is by now excluded at $\sim 6\sigma$, we have to modify our model to make it phenomenologically viable.
Here, $\theta_{13}$ and the mass of the lightest neutrino are
linked~\cite{scaling}, so we need to break $\Z2$ to couple $N_3$ to
the active neutrinos if we want $\theta_{13}\neq 0$. Therefore, a
non-zero $\theta_{13}$ will lead to an unstable DM candidate $\chi$,
with a short lifetime compared to the age of the Universe in general
(see App.~\ref{app:Z2breaking} for an estimate). Note that an explicit (soft) $\Z2$ breaking by the coupling $\overline{N}^c_1 N_3$ does not lead to IH, but rather an $L_e$ symmetric $\mathcal{M}_\nu$~\eqref{eq:symmetricforms}. Correspondingly, the scalar sector needs to be enlarged quite a bit to achieve IH with non-vanishing $\theta_{13}$, which is why we will
not discuss this model any further. Without touching the $\Z2$
symmetry we could of course introduce a Higgs triplet (type-II seesaw)
to generate $\theta_{13}\neq 0$, but once again the scalar sector
blows up. Another solution would be the introduction of additional scalar
doublets---charged under $U(1)'$---which generate off-diagonal mass
terms for the charged leptons and consequently modify the PMNS mixing
matrix. Obviously this once again complicates the scalar sector of the
model and will therefore not be discussed further. 

In the next section we will show that an extension of the fermion sector can easily generate a non-vanishing reactor angle while maintaining a simple scalar sector and the exact $\Z2$ symmetry.

\subsection{Five right-handed neutrinos and \texorpdfstring{$\Z2$}{Z2} symmetry}
\label{sec:fiveplusZ2}

Since the extension by scalars is cumbersome, we seek out a different
solution to generate $\theta_{13}\neq 0$. The anomaly
condition~\eqref{eq:righthandedanomalies} can be fulfilled for five
right-handed neutrinos with $B+3 \Lemt$ charges $+3$, $+3$, $-3$,
$-3$, and $-3$, respectively. To obtain an invertible
$\mathcal{M}_R$---and therefore an approximate $\Lemt$ symmetric
$\mathcal{M}_\nu$---we once again decouple one of the right-handed
neutrinos ($\chi \equiv N_3$) by imposing a $\Z2$ symmetry. Again,
$\chi$ will be our dark matter candidate, to be discussed in
Sec.~\ref{sec:dark_matter}. 

After symmetry breaking with the scalars $H\sim (\vec 1, \vec 2, +1)(0)$ and $S\sim (\vec 1, \vec 1,0)(+6)$ we obtain the mass matrix
\begin{align}
 \mathcal{M}_\nu \simeq - \matrixx{a & b & 0 & 0 \\0 & 0 & c & d \\ 0 & 0 & e & f} \matrixx{ \mathcal{A} & \mathcal{X}\\ \mathcal{X}^T & \mathcal{B}}^{-1} \matrixx{a & 0 & 0\\ b & 0 & 0\\ 0 & c & e\\ 0 & d & f},
\label{eq:fiveplusZ2}
\end{align}
where $\mathcal{X}$ is an arbitrary $2\times 2$ matrix (the gauge invariant mass terms for the right-handed neutrinos) and $\mathcal{A}$, $\mathcal{B}$ are symmetric $2\times 2$ matrices generated by spontaneous $B+3\Lemt$ breaking.
For $c f - e d \neq 0$ there is no massless neutrino $\alpha\, \nu_\mu
+ \beta\, \nu_\tau$, so we have $\theta_{13}\neq 0$ in general. The
solar mixing angle becomes maximal for
$\mathcal{A},\mathcal{B}\rightarrow 0$, so the breaking scale needs to
be close to the bare mass terms to lower $\theta_{12}$. A large
$\theta_{13}$ in agreement with recent results also forbids a too low
breaking scale, meaning that the breaking parameter should be at least $\epsilon \simeq 0.1$ in our minimal model. For the scatter
plots in Fig.~\ref{fig:scatter} (right) we generated random Yukawa
couplings $|(m_D)_{ij}| \leq 1$,
$|(\mathcal{A})_{ij}|,|(\mathcal{B})_{ij}| \leq \epsilon$ and
$|(\mathcal{X})_{ij}|>1$. Except for the approximate $\Lemt$ symmetry
in the limit $\mathcal{A}_{ij},\mathcal{B}_{ij}\ll \mathcal{X}_{m n}$
(and the corresponding inverted hierarchy) there is no further
structure in $\mathcal{M}_\nu$, so we refrain from any analytical
discussion.

We conclude the section by stressing once more that the spontaneously
broken $B+3\Lemt$ symmetry can provide mass matrices for either normal
or inverted hierarchy, just depending on whether the number of
``active" right-handed neutrinos is odd or even, respectively. Since
anomaly cancellation requires an odd number of $N_i$---at least for
physically interesting charge assignments---the decoupling to get IH
needs to be imposed by additional symmetries, which can easily lead to
stable dark matter candidates. While we discussed only the simplest
decoupling symmetry $\mathbb{Z}_2$, one can of course implement a more
elaborate structure on this sub-sector.

\section{Gauge Sector}
\label{sec:gauge_sector}

\begin{figure}[t]
	\begin{center}
		\includegraphics[width=0.48\textwidth]{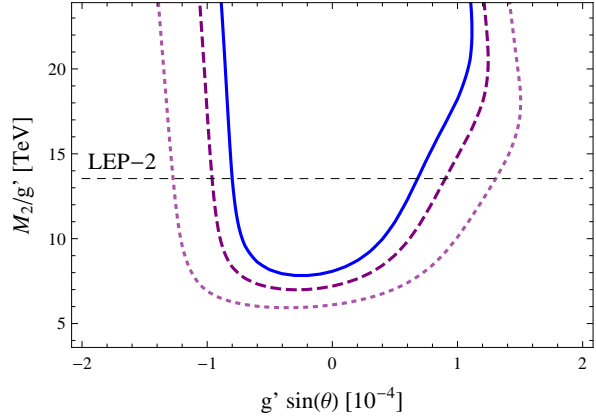}
	\end{center}
		\caption{$\chi^2$ contours ($90\%$, $95\%$ and $99\%$ C.L.) in the $M_2$-$\sin(\theta)$ plane. The horizontal dashed line is the $95\%$ C.L.~lower limit from LEP-2~\cite{LEP-2bounds}.}
	\label{fig:mzpvstheta}
\end{figure}

In this section we will briefly discuss constraints on the neutral
gauge boson of the gauged $B+3\Lemt$ symmetry and possible detection
prospects. The presented results are independent of the neutrino
sector. 
Extending the gauge group of the SM $G_\sm \equiv SU(3)_C \times SU(2)_L \times U(1)_Y$ by $U(1)'$ leads to possible $Z$--$Z'$ mixing, either from the VEV of a scalar in a non-trivial representation of $SU(2)_L \times U(1)_Y$ and $U(1)'$ or via the kinetic mixing angle $\chi$ that connects the $U(1)$ field strength tensors~\cite{kineticmixing}. The relevant Lagrange density $\L = \L_\mathrm{SM} + \L_{Z'} + \L_\mathrm{mix}$ after breaking $SU(2)_L \times U(1)_Y \times U(1)'$ to $U(1)_\mathrm{EM}$ then consists of:
\begin{align}
\begin{split}
	\L_\mathrm{SM} &= -\frac{1}{4} \hat{B}_{\mu\nu} \hat{B}^{\mu\nu} -\frac{1}{4} \hat{W}^a_{\mu\nu} \hat{W}^{a\mu\nu} + \frac{1}{2} \hat{M}_Z^2 \hat{Z}_\mu \hat{Z}^\mu  - \frac{\hat{e}}{\hat{c}_W} j_Y^\mu \hat{B}_\mu -\frac{\hat{e}}{\hat{s}_W} j_{SU(2)}^{a\mu} \hat{W}^a_\mu\,,	\\
	\L_{Z'} &= -\frac{1}{4} \hat{Z}'_{\mu\nu} \hat{Z}'^{\mu\nu}+ \frac{1}{2} \hat{M}_{Z'}^2 \hat{Z}'_\mu \hat{Z}'^\mu - \hat{g}' j'^\mu \hat{Z}'_\mu \,,\\
	\L_\mathrm{mix} &= -\frac{\sin \chi}{2} \hat{Z}'^{\mu\nu}\hat{B}_{\mu\nu} + \delta \hat{M}^2 \hat{Z}'_\mu \hat{Z}^\mu \,.
\end{split}
\label{eq:lagrangian}
\end{align}
Since the above gauge eigenstates have a non-diagonal mass matrix and kinetic terms, the physical mass eigenstates are linear combinations of the hatted fields. Setting for simplicity the kinetic mixing angle $\chi$ to zero, the transformation to the mass eigenstates $Z_1$ and $Z_2$ takes the simple form
\begin{align}
\matrixx{Z_1\\Z_2} = \matrixx{\cos\theta & \sin\theta\\-\sin\theta & \cos \theta} \matrixx{\hat{Z}\\\hat{Z}'} , && \tan 2\theta = \frac{2\, \delta \hat{M}^2}{\hat{M}_Z^2 - \hat{M}_{Z'}^2}\,,
\end{align}
which modifies the couplings of the gauge bosons to fermions (see Ref.~\cite{Langacker:2008yv} for more details). Using a modified version of GAPP~\cite{Erler} to fit our model with an arbitrary scalar sector we obtain the $95\%$ C.L.~limit $|g'\sin\theta|\lesssim 10^{-4}$ (see Fig.~\ref{fig:mzpvstheta}) from electroweak precision data. Constraints for the mass $M_{Z'}$ are obtained from collider searches, as the gauge boson of $U(1)_{B+3\Lemt}$ couples directly to first-generation particles.
LEP-2 searches for new physics give a stronger limit than Tevatron, namely $M_{Z'}/g' \gtrsim \unit[13.5]{TeV}$ at $95\%$ C.L.~\cite{LEP-2bounds}, because the $Z'$ couples strongly to the electron ($Y'(e) = 3$).

In the following we will ignore any $Z$--$Z'$ mixing, be it mass mixing (not induced at tree-level in our minimal model) or kinetic mixing; with $\L_\mathrm{mix}=0$ we can omit all the hats of the parameters in Eq.~\eqref{eq:lagrangian}.

With $n_N$ heavy neutrinos below $M_{Z'}$---and with charges $|Y'(N)|=3$---we can calculate the $Z'$ width:
\begin{align}
\begin{split}
 \Gamma (Z'\ra f \overline{f}) &\simeq \frac{{g'}^2}{12\pi} M_{Z'} \left( 3 {Y'_\nu}^2/2 +n_N {Y'_N}^2/2 + 3 {Y'_\ell}^2 + 3 N_c {Y'_u}^2 + 3 N_c {Y'_d}^2\right) \\
 &= \frac{{g'}^2}{24\pi} M_{Z'} \left( 85 + 27\,n_N \right) ,
\end{split}
\end{align}
with the number of colors $N_c = 3$. 
The main contribution comes from the leptons, because of the large $B+3\Lemt$ charge, which can be used to distinguish this model from the similar $B-L$ model at colliders.
The prospects of detecting the heavy $Z'$ at the LHC were discussed in Ref.~\cite{Lee:2010hf}; for $g'=0.1$ the final stage of the LHC ($\sqrt{s} = \unit[14]{TeV}$, integrated luminosity $ L \simeq \unit[100]{fb^{-1}}$) can probe the model up to $M_{Z'} \simeq \unit[3.6]{TeV}$ via the dilepton $Z'$ resonance.

We note that the non-universal lepton coupling of $B+3 \Lemt$ gives
rise to non-standard neutrino interactions (NSIs), which are usually
parameterized by the non-renormalizable effective Lagrangian
\begin{align}
	\L^\mathrm{eff}_\mathrm{NSI} = -2 \sqrt{2} G_F \epsilon_{\alpha\beta}^{f P} \left[\bar f \gamma^\mu P f \right] \left[ \bar \nu_\alpha \gamma_\mu P_L \nu_\beta\right] ,
	\label{eq:neutralNSI}
\end{align}
in our case obtained upon integrating out the heavy gauge boson $Z'$.
Without going into details, we can estimate
\begin{align}
\epsilon_{\alpha\beta} \sim \frac{v_{EW}^2}{(M_{Z'}/g')^2}\,\diag (1,-1,-1) = \frac{v_{EW}^2}{(M_{Z'}/g')^2}\,\diag (2,0,0) + \frac{v_{EW}^2}{(M_{Z'}/g')^2}\,\diag (1,1,1)\,.
\end{align}
The magnitude is very small ($\epsilon \sim 10^{-4}$) and since the term proportional to the identity matrix does not affect oscillations, we actually only induce $\epsilon_{ee}$, i.e.~modify the usual matter potential, which is hard to measure.

\section{Minimal Scalar Sector}
\label{sec:scalar_sector}

In this section we will discuss the scalar sector of the theory, which
is again independent on the neutrino physics. 
In addition to the usual scalar doublet $H \sim (\vec{1},\vec{2},+1) (0)$ of the SM, we introduce a complex scalar $S\sim (\vec{1},\vec{1},0)(+6)$ that will break the $B+3 \Lemt$ symmetry spontaneously. The discussion is analogous to the highly discussed minimal $B-L$ scalar sector~\cite{minimalB-L}. The potential has the simple form
\begin{align}
	V (H, S) = -\mu_1^2 |H|^2 + \lambda_1 |H|^4 - \mu_2^2 |S|^2 + \lambda_2 |S|^4 + \delta |S|^2 |H|^2\,,
\end{align}
where we assume $\mu_i^2 >0$ to generate VEVs $v \equiv \sqrt{2}|\langle H\rangle |$ and $v_S\equiv \sqrt{2}|\langle S\rangle |$. The positivity of the potential gives the constraints $\lambda_i > 0$ and $	\lambda_1 \lambda_2 > \delta^2/4$.
In unitary gauge the charged component of $H$ is absorbed by $W^\pm$, the pseudoscalar neutral component by $Z$, and the pseudoscalar component of $S$ by $Z'$, hence we may go to the physical basis $H\ra (0, (h+v)/\sqrt{2})^T$, $S\ra (s+v_S)/\sqrt{2}$, which after the replacement of $\mu_i^2$ by the VEVs gives the potential:
\begin{align}
\begin{split}
	V(h,s) &= \lambda_1 v^2 h^2 + \lambda_2 v_S^2 s^2 + \delta v v_S h s \\
	&\quad + \lambda_1 v h^3 + \frac{\lambda_1}{4} h^4 +\lambda_2 v_S s^3 + \frac{\lambda_2}{4} s^4+ \frac{\delta}{4} h^2 s^2 + \frac{\delta}{2} v h s^2 +  \frac{\delta}{2} v_S h^2 s\,.
\end{split}
\end{align}
The resulting mass matrix for the neutral scalars $h$ and $s$
\begin{align}
	\mathcal{M}_\mathrm{scalar}^2 = \matrixx{2 \lambda_1 v^2 & \delta v v_S\\ \delta v v_S & 2 \lambda_2 v_S^2}
	\label{eq:scalar_mass_matrix}
\end{align}
leads to the mass eigenstates $\phi_{1}$ and $\phi_2$:
\begin{align}
	\matrixx{\phi_1\\\phi_2} = \matrixx{\cos \alpha & - \sin \alpha\\ \sin\alpha & \cos \alpha}\matrixx{h\\s}, &&
	\tan 2\alpha = \frac{ \delta v v_S}{ \lambda_2 v_S^2 - \lambda_1 v^2}\,,
\end{align}
with the masses $m_{1,2}^2 = \lambda_1 v^2 + \lambda_2 v_S^2 \mp
\sqrt{(\lambda_2 v_S^2 - \lambda_1 v^2)^2 + \delta^2 v_S^2 v^2}$. In
the limit $v_S \gg v$ we obtain $\alpha \simeq \delta v / 2 \lambda_2
v_S$ and $m_1^2 \simeq 2 (\lambda_1 - \delta^2/4\lambda_2) v^2$, so
the Higgs mass is reduced compared to the SM.  

The LEP-2 bounds on $M_{Z'}/g'$ translate into the constraint $v_S > \unit[2.3]{TeV}$, which is to be compared to the VEV in minimal $B-L$ models $v_{B-L}>3$--$\unit[3.5]{TeV}$. The masses of $Z'$, $\phi_2$ and $\chi$ can of course be smaller, since they involve additional coupling constants (that are completely independent of each other):
\begin{align}
	M_{Z'} = 6 g' v_S\,, &&
	m_2 \simeq m_s \simeq \sqrt{2 \lambda_2} v_S\,, &&
	M_\chi = \sqrt{2}Y_\chi v_S\,.
\label{eq:brokenmasses}
\end{align}
Seeing as the VEV $v_S$ is connected to the seesaw scale (Eqs.~\eqref{eq:seesaw_oom},\eqref{eq:Lemtsymmetric}) one could also consider $v_S \sim \unit[10^{15}]{GeV}$, which would make $Z'$ and $s$ pretty much impossible to observe. We will therefore focus on the low-energy end of the seesaw scale, which can lead to observable effects. 
The arising effects are nearly identical to the highly discussed minimal $B-L$ scalar sector~\cite{minimalB-L}, the main difference being a larger $Z'$ coupling to leptons, right-handed neutrinos and the scalar $s$; for these particles, $B-L$ results can be translated via $g'\ra 3 \, g'$, while the coupling strength to quarks does not change. A future lepton collider would therefore be the ideal machine to test this model and distinguish it from $B-L$ by the decay products of the $Z'$ resonance.

\section{Dark Matter}
\label{sec:dark_matter}

\begin{figure}[t]
	\begin{center}
		\includegraphics[width=0.48\textwidth]{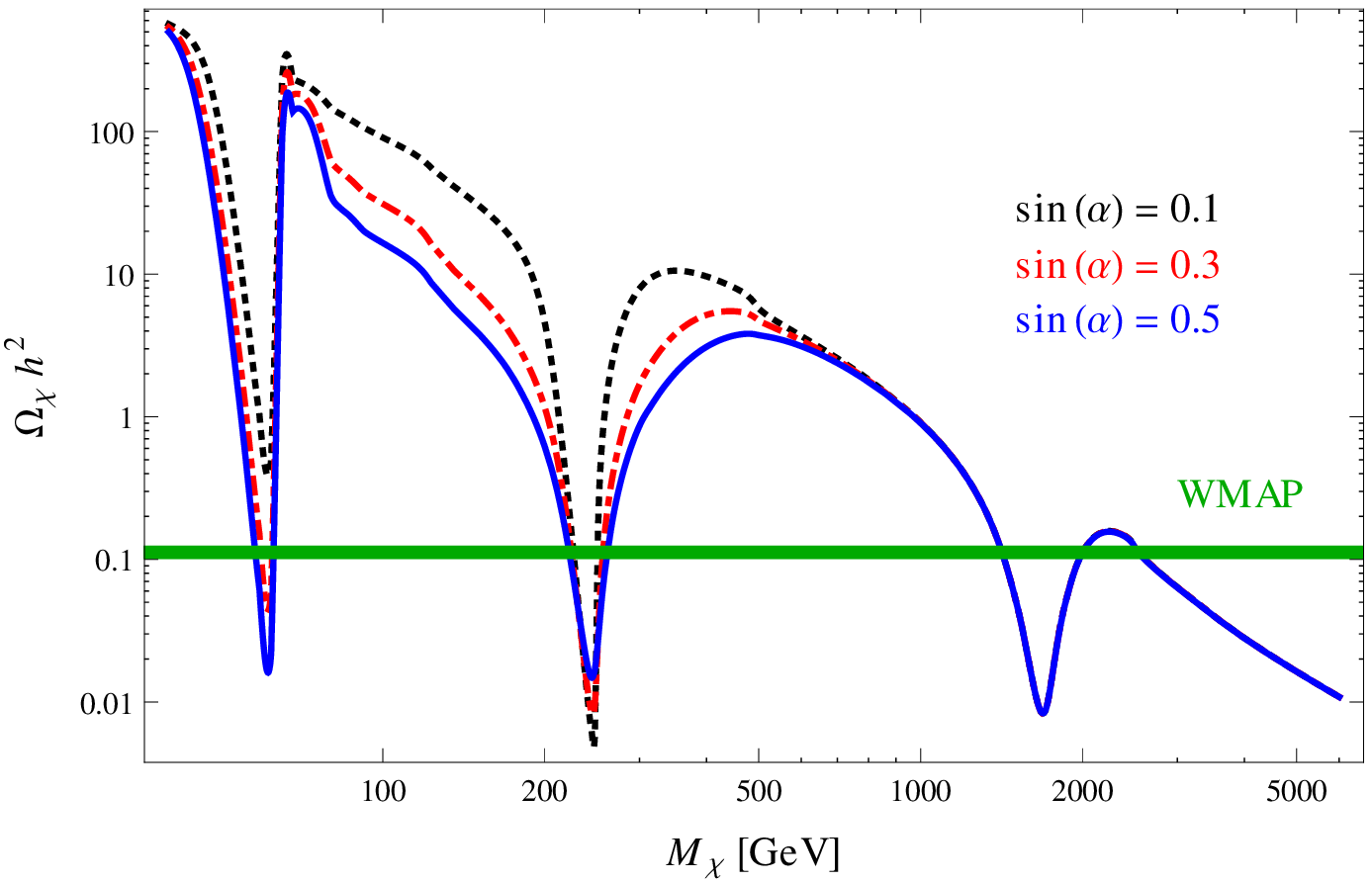}\hspace{1ex}
		\includegraphics[width=0.48\textwidth]{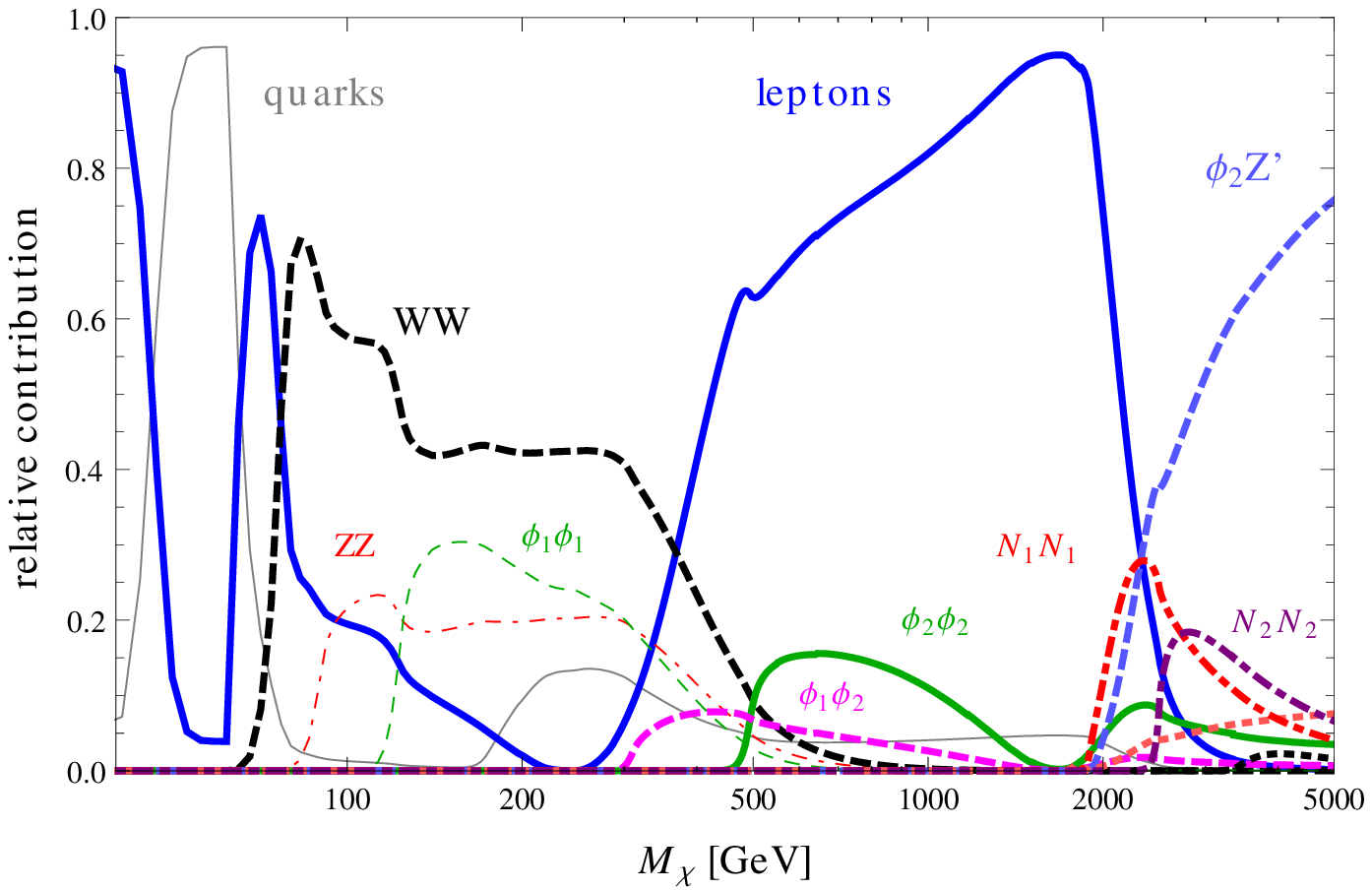}
	\end{center}
		\caption{Left: Relic density of $\chi$ for the
parameters $m_1=\unit[125]{GeV}$, $m_2=\unit[500]{GeV}$,
$v_S=\unit[2.3]{TeV}$, $g'=0.25$, $N_1 = \unit[1.9]{TeV}$, $N_2 =
\unit[2.5]{TeV}$, and $\sin\alpha = 0.5$ (blue), $0.3$ (red) and $0.1$
(black). This puts the $\phi_1$, $\phi_2$ and $Z'$ resonances at $\sim \unit[60]{GeV}$, $\unit[250]{GeV}$ and $\unit[1.7]{TeV}$, respectively. The green band shows the $3\sigma$ range measured by
WMAP. Right: Relative contribution to the relic density by the
processes $\chi \chi \ra q\overline{q}$ (sum over all quarks), leptons
(including neutrinos), $Z Z$, etc., for $\sin\alpha = 0.3$. } 
	\label{fig:relic}
\end{figure}

As we have seen in the previous sections, our model leads to a stable right-handed neutrino $\chi$ which interacts with $Z'$ and the $\phi_i$ via the Lagrangian from Eq.~\eqref{eq:N3lagrangian}.
The measured relic density~\cite{wmap} $\Omega_\chi h^2 = 0.1123 \pm
0.0035$ can be obtained around either of the scalar $s$-channel
resonances $M_\chi \simeq m_i/2$, but for the $\phi_1$-resonance one needs a large
mixing angle $\alpha$. Choosing parameters that make the model
testable at LHC and direct DM detection experiments---$M_\chi \sim
10$--$\unit[100]{GeV}$, $m_2 \sim \unit[100]{GeV}$---can lead to
viable DM relic abundance in complete analogy to
Ref.~\cite{B-Ldarkmatter}, where a $\Z2$ symmetry is added to the
minimal $B-L$ model to make one of the right-handed neutrinos
stable. We stress however that the $\Z2$ in our model was not
introduced to make a particle stable, but to generate the right flavor
symmetry in the neutrino mass matrix. The stability of $\chi$ is in
that sense just a welcome accident.\footnote{Note that we need an exact $\Z2$ for dark matter, while a valid IH solution could also work with a broken $\Z2$. This would however necessitate a more complicated model, so Occam's razor suggests an exact $\Z2$.} We show the relic abundance of
$\chi$ as a function of its mass and the $h$-$s$ mixing angle $\alpha$
in Fig.~\ref{fig:relic}, as calculated with a modified version of
microMEGAs~\cite{micromegas}. There is no difference between the
$B+3\Lemt$ model and the $B-L$ model in the region $M_\chi\ll M_{Z'}$
of parameter space, because the $Z'$ plays a sub-dominant role for the
properties of the scalars, so we refer to Ref.~\cite{B-Ldarkmatter}
for exact formulae of the relevant cross sections and discussions of
direct detection signals etc. Additional work on $B-L$ in connection
with dark matter has been done in Refs.~\cite{moreB-L,SUSYB-L}. 

Values around $M_\chi \sim \unit[100]{GeV}$ are an interesting
limiting case for collider searches. However, since $\chi$, $N_i$, $Z'$, and
$\phi_2$ all obtain their masses from $B+3\Lemt$
breaking~\eqref{eq:brokenmasses}, we naturally would expect their
masses to be similar: 
\begin{align}
 M_{Z'} \sim m_2 \sim M_\chi \sim M_{N_i}\,.
 \label{eq:similarmasses}
\end{align}
To satisfy collider constraints and give a valid seesaw mechanism one needs the scale for these masses to be above $1$--$\unit[10]{TeV}$, but it can of course be even higher. A valid relic density can be obtained yet again around the $\phi_2$ resonance, since we expect $\chi$ and $\phi_2$ to have similar masses anyway. The important annihilation channels are then $\chi\chi \ra $~leptons, $WW$, $ZZ$ and $\phi_1\phi_1$. The latter three have a fixed ratio at the resonance, because for $m_2 \gg m_1,M_Z$ one calculates
\begin{align}
 \Gamma (\phi_2 \ra W^+ W^- ) \simeq  2 \Gamma (\phi_2 \ra Z Z )  \simeq  2 \Gamma (\phi_2 \ra \phi_1 \phi_1 ) \simeq \frac{m_2^3}{16 \pi v^2} \sin^2\alpha\,.
\end{align}
For $M_\chi \simeq m_2/2 > m_t$ there is of course the additional important decay into top quarks.
However, for a DM candidate this heavy, we also have a $Z'$ resonance $M_\chi \simeq M_{Z'}/2$ independent of the mixing angle $\alpha$. Due to the different coupling of our $Z'$ compared to $B-L$, this $Z'$ resonance is particularly interesting to distinguish the models.
The interactions between fermions and $Z'$ are given by
\begin{align}
\begin{split}
 \L \ \supset\  g' Z'_\mu \biggl( 
&-\frac{3}{2}\, \overline{\chi} \gamma^\mu \gamma_5 \chi  
+\frac{1}{3}\, \sum_q \overline{q}\gamma^\mu q  
-3\, \overline{e}\gamma^\mu e 
+3\, \overline{\tau}\gamma^\mu \tau  \\
&  +\frac{3}{2}\,  \overline{\nu}_e \gamma^\mu (-\gamma_5) \nu_e 
-\frac{3}{2}\,  \overline{\nu}_\tau \gamma^\mu (-\gamma_5) \nu_\tau
+\frac{3}{2}\, \overline{N}_1 \gamma^\mu (+\gamma_5)N_1+\ldots \biggr)\,,
\end{split}
\end{align}
where $\chi$ and the neutrinos are written as Majorana fermions. The structure of the effective operators $\overline{\chi} \gamma^\mu \gamma_5 \chi\, \overline{f}\gamma_\mu f$ upon integrating out $Z'$ leads to spin-independent (SI) and spin-dependent (SD) interactions in the non-relativistic limit, suppressed by $v^2$ (velocity) and $q^2$ (momentum transfer), respectively, as discussed in Ref.~\cite{operatoranalysis}.

Around the $Z'$ resonance, the relevant processes $\chi\chi \ra Z'\ra f\overline{f}$ lead to the thermally averaged cross section $\langle \sigma v \rangle \simeq a + b v^2$ with $a=0$ and
\begin{align}
	b \simeq \frac{2 {g'}^4}{3\pi} \frac{M_\chi^2}{(M_{Z'}^2 - 4 M_\chi^2)^2 + \Gamma_{Z'}^2 M_{Z'}^2} \sum_f {Y'}_f^2 {Y'}_\chi^2\,,
\end{align}
where we neglected the fermion masses for simplicity. This can be used to calculate the freeze-out temperature and the relic density $\Omega_\chi h^2 \sim 1/b$~\cite{B-Ldarkmatter} of $\chi$. Due to the larger coupling of $Z'$ to leptons compared to $B-L$, the annihilation channels around the $Z'$ resonance are mainly $\ell \overline{\ell}$, $\nu \overline{\nu}$, and also $N_i N_i$ if $M_{N_i} \lesssim M_{Z'}/2$. 
At this point it matters whether we take $\chi$ from
Sec.~\ref{sec:threeplusZ2} or Sec.~\ref{sec:fiveplusZ2}, because the
models differ in the number of heavy neutrinos. However, additional
right-handed neutrinos do not change the discussion qualitatively, so
we will perform our calculations with $n_N = 3$
(Sec.~\ref{sec:threeplusZ2}) for simplicity. In Fig.~\ref{fig:relic}
we already showed the relic density of $\chi$ and the contributing
processes around the $Z'$ resonance.

\begin{figure}[tb]
	\begin{center}
		\includegraphics[width=0.48\textwidth]{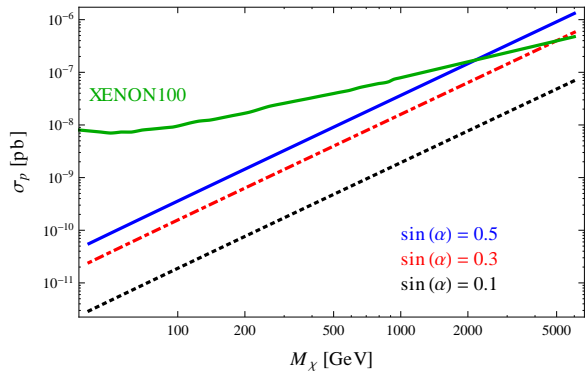}
	\end{center}
		\caption{Spin-independent cross section of $\chi$ with a proton with the same parameters as Fig.~\ref{fig:relic}, i.e.~$m_1=\unit[125]{GeV}$, $m_2=\unit[500]{GeV}$, $v_S=\unit[2.3]{TeV}$, $g'=0.25$, $N_1 = \unit[1.9]{TeV}$, $N_2 = \unit[2.5]{TeV}$, and $\sin\alpha = 0.5$ (blue), $0.3$ (red, dashed) and $0.1$ (black, dotted). Also shown is the XENON100 $90\%$~C.L.~exclusion from Ref.~\cite{xenon100}.}
	\label{fig:directdetection}
\end{figure}

While it is clear from Fig.~\ref{fig:relic} that the $Z'$ channel can lead to the proper relic density (even for $\sin\alpha = 0$), direct detection signals from $Z'$
interactions are difficult to measure due to the Lorentz structure of the effective operator $\overline{\chi} \gamma^\mu \gamma_5 \chi\, \overline{f}\gamma_\mu f$. Since direct detection
occurs via $t$-channel $Z'$~exchange, there is no resonance boost like
in the annihilation case. The SD operators $\overline{\chi} \gamma^\mu
\gamma_5 \chi\, \overline{f}\gamma_\mu \gamma_5 f$---which do not
suffer from $q^2$ or $v^2$ suppression---can only be obtained via
electroweak loops or $Z$--$Z'$ mixing, which once again suppresses
them. Correspondingly, direct detection experiments will not be
sensitive to $Z'$ exchange, so the cross section will be dominated by the
scalar-induced operator $\overline\chi \chi\, \overline q q$, which
gives SI cross sections proportional to $\sin^2 2\alpha  \,
M_\chi^2/v_S^2$. We show the cross sections for $\chi p\ra \chi p$ in
Fig.~\ref{fig:directdetection} (as calculated with microMEGAs) for the same parameters as in
Fig.~\ref{fig:relic}. The right relic density can be obtained for
example at the $\phi_2$ resonance with $M_\chi \simeq
\unit[225]{GeV}$, which gives a cross section $\sigma_p/\sin^2 2\alpha
\simeq \unit[2.5\times 10^{-9}]{pb}$. This evades current XENON100
bounds~\cite{xenon100} but can be probed in future experiments like
XENON1T~\cite{B-Ldarkmatter}.

We note that a supersymmetric extension of this model might result in
$\alpha \ll 1$---making the $Z'$ resonance crucial for relic
abundance---similar to a supersymmetric extension of the $B-L$ model
of Ref.~\cite{B-Ldarkmatter} discussed in Ref.~\cite{SUSYB-L}. 

In any case, dark matter experiments will not be able to distinguish
between the various $B-\sum_\ell x_\ell L_\ell$ models. This has to be
done in collider experiments or with precision observables such as
magnetic moments.

\section{Conclusion}
\label{sec:conclusion}

We presented the minimal model for a local $B+ 3 (L_e - L_\mu -
L_\tau)$ symmetry. Direct detection limits demand a breaking scale of
at least $1$--$\unit[10]{TeV}$ and to cancel anomalies we need to
introduce right-handed neutrinos. Correspondingly, we identify the
breaking scale with the seesaw scale (in fact, slightly below) and
obtain low energy neutrino mass matrices that approximately conserve
$L_e$ or $L_e - L_\mu - L_\tau$, which are the championed symmetries
behind normal and inverted hierarchy, respectively. The latter can be
obtained if a $\Z2$ symmetry is added to the model, resulting in a
stable, heavy right-handed neutrino which serves as dark matter. We
stress that the $\Z2$ is introduced to obtain the flavor structure
associated with the inverted hierarchy, the DM stability is somewhat accidental. The heavy gauge boson $Z'$ and the leftover scalar from spontaneous $B+ 3 (L_e - L_\mu - L_\tau)$ breaking are the only mediators to the DM sector and are in principle observable at the LHC.
Depending on the number $n_N$ of right-handed neutrinos, our model can produce $\theta_{13}=0$ ($n_N=3$ plus $\Z2$) or $\theta_{13}\neq 0$ ($n_N\geq 5$ (odd) plus $\Z2$), making $\theta_{13}$ a parameter to test the number of right-handed neutrinos in our model.

The dark matter candidate $\chi$ interacts with the Standard Model via scalar mixing and $Z'$; the measured relic abundance can be obtained around any of the $s$-channel resonances $M_\chi \simeq m_i/2$, $M_{Z'}/2$. The $Z'$~contribution to direct detection measurements is highly suppressed due to the Lorentz-structure of the effective operator $\overline\chi \gamma^\mu \gamma_5 \chi \, \overline f \gamma_\mu f$, so direct detection cross sections are dominated by scalar exchange and can be probed in future experiments.

Many of the phenomenological aspects of our models are very similar to previously discussed $B-L$ analyses. However, the fact that our modified gauge group includes flavor information, makes it possible to provide predictions on neutrino mixing and mass spectrum, which is impossible in theories based on $B-L$.

\begin{acknowledgments}
This work was supported by the ERC under the Starting Grant 
MANITOP. J.H.~acknowledges support by the IMPRS for Precision Tests of
Fundamental Symmetries and thanks the theory group of Kanazawa University, where part of this
work was performed, for very kind hospitality. 
\end{acknowledgments} 

\appendix

\section{Triangle Anomalies}
\label{app:anomalies}

We introduce $n_N$ right-handed neutrinos $N_i$ with $U(1)_{B- x_e L_e - x_\mu L_\mu - x_\tau L_\tau}$ quantum numbers $Y' (N_i)$. The gauge group representations of the first-generation fermions are shown in Tab.~\ref{tab:quantum_numbers2}, for the second and third generation $x_e$ has to be replaced by $x_\mu$ and $x_\tau$ respectively.
\begin{table}[ht]
\centering
\begin{tabular}[t]{|l|l|l|}
\hline                               
     $L_e = \matrixx{\nu\\e}_L \sim (\vec{1},\vec{2},-1)(-x_e)$ & $e_R^c \sim (\vec{1},\vec{1},+2)(x_e)$ & $N_i^c \sim (\vec{1}, \vec{1},0)(Y' (N_i^c))$\\
     \hline
     $Q^u_L = \matrixx{u\\ d}_L \sim (\vec{3},\vec{2},+\frac{1}{3})(+\frac{1}{3})$ & $u_R^c \sim (\vec{\overline{3}},\vec{1},-\frac{4}{3})(-\frac{1}{3})$ & $d_R^c \sim (\vec{\overline{3}},\vec{1},+\frac{2}{3})(-\frac{1}{3})$ \\
\hline 
\end{tabular}
\caption{$SU(3)_C\times SU(2)_L\times U(1)_Y \times U(1)_{B- x_e L_e - x_\mu L_\mu - x_\tau L_\tau}$ representations of left-handed SM fermions (only first generation shown) and right-handed neutrinos $N_i$.}
\label{tab:quantum_numbers2}
\end{table}

Defining for simplicity $Y' \equiv B- x_e L_e - x_\mu L_\mu - x_\tau L_\tau$ and $U(1)' \equiv U(1)_{Y'}$ we can calculate the triangle anomalies of the model~\cite{B-stuff,Geng:1988pr}:
\begin{align}
\begin{split}
	U(1)' &- \mathrm{grav} - \mathrm{grav}: 
\sum Y' = N_g N_c \left( 2\, \left(\frac{1}{3}\right) - \left(\frac{1}{3}\right)  -  \left(\frac{1}{3}\right) \right) +\sum_\ell \left( x_\ell -2\cdot x_\ell\right) + \sum_i Y' (N_i^c) \\
&\hspace{3.45cm}= \sum_i Y' (N_i^c) - \sum_\ell x_\ell\,,\\
	U(1)' &-U(1)'  -U(1)' : 
\sum {Y'}^3 =  \sum_i Y'^3 (N_i^c) - \sum_\ell x_\ell^3\,,\\
	U(1)' &-U(1)'  -U(1)_Y : 
\sum Y'^2 Y = N_g N_c \left( 2\, \left(\frac{1}{3}\right)^2 \, \left(\frac{1}{3}\right) + \left(-\frac{1}{3}\right)^2 \left(- \frac{4}{3}\right)+ \left(-\frac{1}{3}\right)^2 \left( \frac{2}{3}\right)\right) = 0\,,\\
	U(1)' &-U(1)_Y  -U(1)_Y : 
\sum Y' Y^2 = N_g N_c \left( 2\, \left(\frac{1}{3}\right) \, \left(\frac{1}{3}\right)^2 + \left(-\frac{1}{3}\right) \left(- \frac{4}{3}\right)^2+ \left(-\frac{1}{3}\right) \left( \frac{2}{3}\right)^2 \right)\\
 &\hspace{4.5cm}+ \sum_\ell (2^2 x_\ell - 2 x_\ell) =-2 N_g N_c  \left(\frac{1}{3}\right) +2 \sum_\ell  x_\ell \,,\\
	U(1)' &- SU(3) - SU(3) : 
\sum_{\vec{3},\vec{\overline{3}}} Y' =  N_g N_c \left( 2\, \left(\frac{1}{3}\right) - \left(\frac{1}{3}\right)  -  \left(\frac{1}{3}\right) \right) = 0\,,\\
	U(1)' &- SU(2) - SU(2): 
\sum_\vec{2} Y' = 2\, N_g N_c \left(\frac{1}{3}\right) -2  \sum_\ell x_\ell\,,
\end{split}
\end{align}
where we introduced the number of generations $N_g = 3$ and the number of colors $N_c=3$. Thus the conditions for anomaly-freedom are
\begin{align}
  \sum_{i}^{n_N} Y' (N_i^c) =  \sum_\ell x_\ell=3\,, && \text{ and } &&
  \sum_i Y'^3 (N_i^c) = \sum_\ell x_\ell^3\,.
\end{align}
In this paper we discuss the choice $x_e = -x_\mu = - x_\tau = -3$, which leads to the conditions for the right-handed neutrino charges
\begin{align}
  \sum_{i}^{n_N} Y' (N_i) =-3\,, && \text{ and } &&
  \sum_i Y'^3 (N_i) = -3^3\,.
\end{align}

\section{Unstable dark matter from \texorpdfstring{$\Z2$}{Z2} Breaking}
\label{app:Z2breaking}

We will discuss the connection between the lifetime of $\chi \equiv N_3$ and $\theta_{13}$ in the model of Sec.~\ref{sec:threeplusZ2}, as they are both connected to $\Z2$ breaking. $\Z2$ breaking in $\mathcal{M}_R$ does not induce a non-zero $\theta_{13}$, so we assume there are breaking terms in $m_D$:\footnote{We do not specify the origin of these parameters; they can be obtained with an additional Higgs doublet or as effective operators from $\Z2$ breaking at the Planck scale. In the former case the additional scalars might contribute to the width of $\chi$.}
\begin{align}
 \L \supset \delta \,\overline{\nu}_\mu \chi + \epsilon \,\overline{\nu}_\tau \chi +\hc\,,
\label{eq:Z2breaking}
\end{align}
where $\delta$ and $\epsilon$ carry no $B+3\Lemt$ charge, wlog. The mass matrix $\mathcal{M}_\nu$ from Eq.~\eqref{eq:Lemtsymmetric} gets perturbed by
\begin{align}
 \Delta \mathcal{M} = - \matrixx{0 & 0 & 0 \\ \cdot & \delta^2/M_\chi & \delta \epsilon/M_\chi \\ \cdot & \cdot & \epsilon^2/M_\chi}.
\end{align}
The lowest neutrino mass is no longer zero but rather $\simeq (\delta-\epsilon)^2/2 M_\chi$, and furthermore
\begin{align}
 \sin \theta_{13} \simeq \frac{\delta^2-\epsilon^2}{M_\chi} \,\frac{R}{\sqrt{18}} \,\frac{1}{\sqrt{|\Delta m_{31}^2|}}\,,
\end{align}
with $R\equiv \Delta m_{21}^2/\Delta m_{31}^2 \simeq 0.03$.
\begin{figure}[bt]
	\begin{center}
		\includegraphics[scale=0.6]{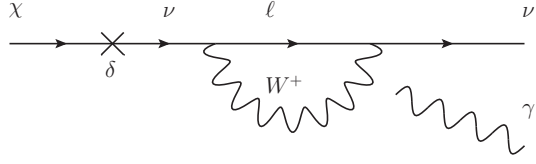}
	\end{center}
		\caption{Decay of the dark matter candidate $\chi$ due to small $Z_2$ breaking $\epsilon$ in the Dirac mass matrix $m_D$.}
	\label{fig:neutrino_decay}
\end{figure}

The Lagrangian~\eqref{eq:Z2breaking} generates the $\chi$ decay via $W$-loop (Fig.~\ref{fig:neutrino_decay}) and tree-level $\chi \ra \nu Z\ra 3\nu$, the latter of which dominates and can be estimated via
\begin{align}
 \Gamma (\chi\ra 3 \nu) \simeq \frac{G_F^2 }{192 \pi^3} \left(\frac{\delta}{M_\chi}\right)^2 \, M_\chi^5\,,
\end{align}
where we set $\epsilon=0$ for simplicity. The decay width is therefore linear in $\sin \theta_{13}$:
\begin{align}
\begin{split}
 \Gamma (\chi\ra 3 \nu) &\simeq \sqrt{18} \,\frac{G_F^2 }{192 \pi^3} \frac{\sqrt{|\Delta m_{31}^2|}}{R} M_\chi^4 \sin \theta_{13}\\
& \simeq \sin\theta_{13}\,\left( \frac{M_\chi}{\unit[100]{GeV}}\right)^4\,\unit[10^{-14}]{GeV}\,,
\end{split}
\label{eq:width13}
\end{align}
resulting in a lifetime compared to the age of the universe
\begin{align}
 \tau_\chi / \tau_\mathrm{Universe} \simeq \frac{10^{-28}}{\sin\theta_{13}} \left( \frac{\unit[100]{GeV}}{M_\chi}\right)^4\,.
\end{align}
Even for the smallest currently allowed $\sin\theta_{13}\simeq 0.03$ (at $3\sigma$) it is not possible to make $\chi$ a sufficiently long-lived cold dark matter candidate.

\section{Majorana Interactions}
\label{app:majorana}

Here we provide some details on the rewriting of the Lagrangian of the
right-handed Dirac fermion $N_3$ in terms of a Majorana fermion
$\chi$. After spontaneous breaking of $B+3\Lemt$ we have 
\begin{align}
\begin{split}
	\L_{N_3} &= i \overline{N_3} \gamma^\mu \left( \del_\mu - i (-3)  g' Z'_\mu\right) P_R N_3 - Y_\chi S\  \overline{N_3}^c P_R N_3- Y_\chi \overline{S}\  \overline{N_3} P_L N_3^c\\
&= i \overline{N_3} \gamma^\mu \left( \del_\mu - i (-3)  g' Z'_\mu\right) P_R N_3 - Y_\chi \frac{v_S}{\sqrt{2}}\ \left( \overline{N_3}^c P_R N_3 +\overline{N_3} P_L N_3^c\right) \left( 1+ \frac{s}{v_S}\right)\,.
\end{split}
\end{align}
Introducing two Majorana fields $\phi = \phi^c$ and $\chi = \chi^c$ so that $N_3 = P_L \phi + P_R \chi$ gives
\begin{align}
 \L_\chi = \frac{i}{2} \chi^T \mathcal{C} \gamma^\mu  \del_\mu \chi
-\frac{3}{2} g' Z'_\mu \, \chi^T \mathcal{C}  \gamma^\mu \gamma_5 \chi 
- Y_\chi \frac{v_S}{\sqrt{2}}\ \chi^T \mathcal{C} \chi \left( 1+ \frac{s}{v_S}\right)\,,
\end{align}
where we omitted the non-interacting $\phi$ and used the Majorana identities\footnote{It proves convenient to add a total derivative to the Lagrangian to replace the kinetic term $\overline{N_3} \slashed{\del} N_3$ by $\frac{1}{2}(\overline{N_3} \slashed{\del} N_3 - (\del^\mu \overline{N_3}) \gamma_\mu N_3)$.}
\begin{align}
 \overline{\chi} \gamma^\mu \chi = 0 = \overline{\chi} \gamma_\mu \gamma^5 \del^\mu \chi - (\del^\mu \overline{\chi}) \gamma_\mu \gamma^5 \chi\,,
\end{align}
as well as the general equations $[\mathcal{C}, \gamma_5]=0$, $\overline{\mathcal{C}} = \mathcal{C}$ and $\overline{P}_{L,R} = P_{R,L}$. Due to the Majorana condition we also have $\chi^T \mathcal{C} = \overline{\chi}$.
Reading off the mass $m_\chi/2 =  Y_\chi v_S/\sqrt{2}$ we note that $\chi$ is to be treated as a real field when computing functional derivatives (additional factor of $2$). 

A similar analysis can be performed for the neutrinos (active and right-handed), with the additional complication of mixing (via seesaw). The only important part is actually the $\gamma_5$ in the $Z'$ and $Z$ interactions, which leads to spin-dependent scattering. For the active neutrinos the $\gamma_5$ stems from a left-handed projector, so we end up with
\begin{align}
\begin{split}
 \L \ \supset \ &+\frac{1}{2} (+3 g') Z'_\mu \, \overline{\nu}_e \gamma^\mu (-\gamma_5) \nu_e +\frac{1}{2} (-3 g') Z'_\mu \, \overline{\nu}_\tau \gamma^\mu (-\gamma_5) \nu_\tau \\
&+\frac{1}{2} (+3 g') Z'_\mu \, \overline{N}_1 \gamma^\mu (+\gamma_5) N_1+\frac{1}{2} (-3 g') Z'_\mu \, \overline{N}_2 \gamma^\mu (+\gamma_5) N_2\,,
\end{split}
\end{align}
etc., where all fermions are Majorana particles.


\begin{thebibliography}{99}

  
\bibitem{Choubey:2004hn}
  S.~Choubey and W.~Rodejohann,
  Eur.\ Phys.\ J.\  C {\bf 40}, 259 (2005)
  [\href{http://arxiv.org/abs/hep-ph/0411190}{arXiv:hep-ph/0411190}].


\bibitem{zero}
  R.~Foot,
  Mod.\ Phys.\ Lett.\  A {\bf 6}, 527 (1991);\\
  X.~G.~He, G.~C.~Joshi, H.~Lew and R.~R.~Volkas,
  Phys.\ Rev.\  D {\bf 44}, 2118 (1991);\\
  R.~Foot, X.~G.~He, H.~Lew and R.~R.~Volkas,
  Phys.\ Rev.\  D {\bf 50}, 4571 (1994)
  [\href{http://arxiv.org/abs/hep-ph/9401250}{arXiv:hep-ph/9401250}].


\bibitem{mu-tau-short}
For a collection of references see
  J.~Heeck and W.~Rodejohann,
  Phys.\ Rev.\  D {\bf 84}, 075007 (2011)
  [\href{http://arxiv.org/abs/1107.5238}{arXiv:1107.5238} [hep-ph]].


\bibitem{Lemt}
  S.~T.~Petcov,
  Phys.\ Lett.\  B {\bf 110}, 245 (1982);\\
  C.~N.~Leung and S.~T.~Petcov,
  Phys.\ Lett.\  B {\bf 125}, 461 (1983);\\
  G.~C.~Branco, W.~Grimus and L.~Lavoura,
  Nucl.\ Phys.\  B {\bf 312}, 492 (1989);\\
  A.~S.~Joshipura and S.~D.~Rindani,
  Eur.\ Phys.\ J.\  C {\bf 14}, 85 (2000)
  [\href{http://arxiv.org/abs/hep-ph/9811252}{arXiv:hep-ph/9811252}];\\
  R.~N.~Mohapatra, A.~Perez-Lorenzana and C.~A.~de Sousa Pires,
  Phys.\ Lett.\  B {\bf 474}, 355 (2000)
  [\href{http://arxiv.org/abs/hep-ph/9911395}{arXiv:hep-ph/9911395}];\\
  L.~Lavoura,
  Phys.\ Rev.\  D {\bf 62}, 093011 (2000)
  [\href{http://arxiv.org/abs/hep-ph/0005321}{arXiv:hep-ph/0005321}];\\
  K.~S.~Babu and R.~N.~Mohapatra,
  Phys.\ Lett.\  B {\bf 532}, 77 (2002)
  [\href{http://arxiv.org/abs/hep-ph/0201176}{arXiv:hep-ph/0201176}];\\
  H.~J.~He, D.~A.~Dicus and J.~N.~Ng,
  Phys.\ Lett.\  B {\bf 536}, 83 (2002)
  [\href{http://arxiv.org/abs/hep-ph/0203237}{arXiv:hep-ph/0203237}];\\
  S.~T.~Petcov and W.~Rodejohann,
  Phys.\ Rev.\  D {\bf 71}, 073002 (2005)
  [\href{http://arxiv.org/abs/hep-ph/0409135}{arXiv:hep-ph/0409135}];\\
  G.~Altarelli and R.~Franceschini,
  JHEP {\bf 0603}, 047 (2006)
  [\href{http://arxiv.org/abs/hep-ph/0512202}{arXiv:hep-ph/0512202}];\\
  D.~Meloni,
  JHEP {\bf 1202}, 090 (2012)
  [\href{http://arxiv.org/abs/1110.5210}{arXiv:1110.5210} [hep-ph]].


\bibitem{lowLe}
  R.~Barbieri, L.~J.~Hall, D.~Tucker-Smith, A.~Strumia and N.~Weiner,
  JHEP {\bf 9812}, 017 (1998)
  [\href{http://arxiv.org/abs/hep-ph/9807235}{arXiv:hep-ph/9807235}];\\
  W.~Grimus and L.~Lavoura,
  JHEP {\bf 0107}, 045 (2001)
  [\href{http://arxiv.org/abs/hep-ph/0105212}{arXiv:hep-ph/0105212}].


\bibitem{Lemt2}
  L.~Lavoura and W.~Grimus,
  JHEP {\bf 0009}, 007 (2000)
  [\href{http://arxiv.org/abs/hep-ph/0008020}{arXiv:hep-ph/0008020}];\\
  W.~Grimus and L.~Lavoura,
  J.\ Phys.\ G {\bf 31}, 683 (2005)
  [\href{http://arxiv.org/abs/hep-ph/0410279}{arXiv:hep-ph/0410279}].


\bibitem{anomalousU1} 
  Q.~Shafi and Z.~Tavartkiladze,
  Phys.\ Lett.\ B {\bf 482}, 145 (2000)
  [\href{http://arxiv.org/abs/hep-ph/0002150}{arXiv:hep-ph/0002150}];\\
  K.~S.~Babu, A.~G.~Bachri and Z.~Tavartkiladze,
  Int.\ J.\ Mod.\ Phys.\ A {\bf 23}, 1679 (2008)
  [\href{http://arxiv.org/abs/0705.4419}{arXiv:0705.4419} [hep-ph]].


\bibitem{B-3Le}
  E.~Ma, D.~P.~Roy and U.~Sarkar,
  Phys.\ Lett.\  B {\bf 444}, 391 (1998)
  [\href{http://arxiv.org/abs/hep-ph/9810309}{arXiv:hep-ph/9810309}].


\bibitem{B-stuff}
  E.~Ma,
  Phys.\ Lett.\  B {\bf 433}, 74 (1998)
  [\href{http://arxiv.org/abs/hep-ph/9709474}{arXiv:hep-ph/9709474}];\\
  E.~Ma and D.~P.~Roy,
  Phys.\ Rev.\  D {\bf 58}, 095005 (1998)
  [\href{http://arxiv.org/abs/hep-ph/9806210}{arXiv:hep-ph/9806210}];\\
  E.~Ma and U.~Sarkar,
  Phys.\ Lett.\  B {\bf 439}, 95 (1998)
  [\href{http://arxiv.org/abs/hep-ph/9807307}{arXiv:hep-ph/9807307}];\\
  E.~Ma and D.~P.~Roy,
  Phys.\ Rev.\  D {\bf 59}, 097702 (1999)
  [\href{http://arxiv.org/abs/hep-ph/9811266}{arXiv:hep-ph/9811266}].


\bibitem{Chang:2000xy}
  L.~N.~Chang, O.~Lebedev, W.~Loinaz and T.~Takeuchi,
  Phys.\ Rev.\  D {\bf 63}, 074013 (2001)
  [\href{http://arxiv.org/abs/hep-ph/0010118}{arXiv:hep-ph/0010118}].


  \bibitem{Davoudiasl:2011sz} 
  H.~Davoudiasl, H.~S.~Lee and W.~J.~Marciano,
  Phys.\ Rev.\ D {\bf 84}, 013009 (2011)
  [\href{http://arxiv.org/abs/1102.5352}{arXiv:1102.5352} [hep-ph]].


\bibitem{Salvioni:2009jp} 
  E.~Salvioni, A.~Strumia, G.~Villadoro and F.~Zwirner,
  JHEP {\bf 1003}, 010 (2010)
  [\href{http://arxiv.org/abs/0911.1450}{arXiv:0911.1450} [hep-ph]].


\bibitem{Chen:2011de} 
  M.~C.~Chen and J.~Huang,
  Mod.\ Phys.\ Lett.\ A {\bf 26}, 1147 (2011)
  [\href{http://arxiv.org/abs/1105.3188}{arXiv:1105.3188} [hep-ph]].


\bibitem{Lee:2010hf}
  H.~S.~Lee and E.~Ma,
  Phys.\ Lett.\  B {\bf 688}, 319 (2010)
  [\href{http://arxiv.org/abs/1001.0768}{arXiv:1001.0768} [hep-ph]].


\bibitem{moreSUSYB-L}
  S.~Khalil and A.~Masiero,
  Phys.\ Lett.\ B {\bf 665}, 374 (2008)
  [\href{http://arxiv.org/abs/0710.3525}{arXiv:0710.3525} [hep-ph]];\\
  P.~Fileviez Perez and S.~Spinner,
  Phys.\ Rev.\ D {\bf 83}, 035004 (2011)
  [\href{http://arxiv.org/abs/1005.4930}{arXiv:1005.4930} [hep-ph]];\\
  B.~O'Leary, W.~Porod and F.~Staub,
  JHEP {\bf 1205}, 042 (2012)
  [\href{http://arxiv.org/abs/1112.4600}{arXiv:1112.4600} [hep-ph]].


\bibitem{SUSYB-L}
  Z.~M.~Burell and N.~Okada,
  Phys.\ Rev.\  D {\bf 85}, 055011 (2012)
  [\href{http://arxiv.org/abs/1111.1789}{arXiv:1111.1789} [hep-ph]].


\bibitem{hiddenO2} 
  J.~Heeck and W.~Rodejohann,
  JHEP {\bf 1202}, 094 (2012)
  [\href{http://arxiv.org/abs/1112.3628}{arXiv:1112.3628} [hep-ph]].


\bibitem{theta13}
  T.~Schwetz, M.~Tortola and J.~W.~F.~Valle,
  New J.\ Phys.\  {\bf 13}, 063004 (2011)
  [\href{http://arxiv.org/abs/1103.0734}{arXiv:1103.0734} [hep-ph]];\\
  T.~Schwetz, M.~Tortola and J.~W.~F.~Valle,
  New J.\ Phys.\  {\bf 13}, 109401 (2011)
  [\href{http://arxiv.org/abs/1108.1376}{arXiv:1108.1376} [hep-ph]];\\
  G.~L.~Fogli, E.~Lisi, A.~Marrone, A.~Palazzo and A.~M.~Rotunno,
  Phys.\ Rev.\  D {\bf 84}, 053007 (2011)
  [\href{http://arxiv.org/abs/1106.6028}{arXiv:1106.6028} [hep-ph]].


\bibitem{t2k}
  K.~Abe {\it et al.} [T2K Collaboration],
  Phys.\ Rev.\ Lett.\  {\bf 107}, 041801 (2011)
  [\href{http://arxiv.org/abs/1106.2822}{arXiv:1106.2822} [hep-ex]].


\bibitem{doublechooz}
  Y.~Abe {\it et al.} [Double Chooz Collaboration],
  Phys.\ Rev.\ Lett.\  {\bf 108}, 131801 (2012)
  [\href{http://arxiv.org/abs/1112.6353}{arXiv:1112.6353} [hep-ex]].


\bibitem{dayabay} 
  F.~P.~An {\it et al.} [Daya Bay Collaboration],
  Phys.\ Rev.\ Lett.\  {\bf 108}, 171803 (2012)
  [\href{http://arxiv.org/abs/1203.1669}{arXiv:1203.1669} [hep-ex]].


\bibitem{reno} 
  J.~K.~Ahn {\it et al.} [RENO Collaboration],
  Phys.\ Rev.\ Lett.\  {\bf 108}, 191802 (2012)
  [\href{http://arxiv.org/abs/1204.0626}{arXiv:1204.0626} [hep-ex]].


\bibitem{Goswami:2008rt}
  S.~Goswami and A.~Watanabe,
  Phys.\ Rev.\  D {\bf 79}, 033004 (2009)
  [\href{http://arxiv.org/abs/0807.3438}{arXiv:0807.3438} [hep-ph]].


\bibitem{scaling}  
  R.~N.~Mohapatra and W.~Rodejohann,
  Phys.\ Lett.\  B {\bf 644}, 59 (2007)
  [\href{http://arxiv.org/abs/hep-ph/0608111}{arXiv:hep-ph/0608111}];\\
  A.~Blum, R.~N.~Mohapatra and W.~Rodejohann,
  Phys.\ Rev.\  D {\bf 76}, 053003 (2007)
  [\href{http://arxiv.org/abs/0706.3801}{arXiv:0706.3801} [hep-ph]].


\bibitem{Ray:2010fa}
  S.~Ray, W.~Rodejohann and M.~A.~Schmidt,
  Phys.\ Rev.\  D {\bf 83}, 033002 (2011)
  [\href{http://arxiv.org/abs/1010.1206}{arXiv:1010.1206} [hep-ph]].


\bibitem{kineticmixing} 
  B.~Holdom,
  Phys.\ Lett.\ B {\bf 166}, 196 (1986);
  Phys.\ Lett.\ B {\bf 259}, 329 (1991).


\bibitem{Langacker:2008yv} 
  P.~Langacker,
  Rev.\ Mod.\ Phys.\  {\bf 81}, 1199 (2009)
  [\href{http://arxiv.org/abs/0801.1345}{arXiv:0801.1345} [hep-ph]].


\bibitem{Erler}
  J.~Erler, P.~Langacker,
  Phys.\ Lett.\  B {\bf 456}, 68-76 (1999)
  [\href{http://arxiv.org/abs/hep-ph/9903476}{arXiv:hep-ph/9903476}];\\
  J.~Erler,
  \href{http://arxiv.org/abs/hep-ph/0005084}{arXiv:hep-ph/0005084}.


\bibitem{LEP-2bounds}
  [The LEP Collaborations: ALEPH Collaboration, DELPHI Collaboration, L3 Collaboration,
OPAL Collaboration, the LEP Electroweak Working Group, the SLD Electroweak, Heavy
Flavour Groups],
  \href{http://arxiv.org/abs/hep-ex/0312023}{arXiv:hep-ex/0312023};\\
  M.~S.~Carena, A.~Daleo, B.~A.~Dobrescu and T.~M.~P.~Tait,
  Phys.\ Rev.\  D {\bf 70}, 093009 (2004)
  [\href{http://arxiv.org/abs/hep-ph/0408098}{arXiv:hep-ph/0408098}].


\bibitem{minimalB-L}
  L.~Basso, A.~Belyaev, S.~Moretti and C.~H.~Shepherd-Themistocleous,
  Phys.\ Rev.\  D {\bf 80}, 055030 (2009)
  [\href{http://arxiv.org/abs/0812.4313}{arXiv:0812.4313} [hep-ph]];\\
  L.~Basso, S.~Moretti and G.~M.~Pruna,
  Phys.\ Rev.\  D {\bf 83}, 055014 (2011)
  [\href{http://arxiv.org/abs/1011.2612}{arXiv:1011.2612} [hep-ph]].


\bibitem{wmap}
  E.~Komatsu {\it et al.} [WMAP Collaboration],
  Astrophys.\ J.\ Suppl.\  {\bf 192}, 18 (2011)
  [\href{http://arxiv.org/abs/1001.4538}{arXiv:1001.4538} [astro-ph.CO]];\\
  D.~Larson {\it et al.},
  Astrophys.\ J.\ Suppl.\  {\bf 192}, 16 (2011)
  [\href{http://arxiv.org/abs/1001.4635}{arXiv:1001.4635} [astro-ph.CO]].


\bibitem{B-Ldarkmatter}
  N.~Okada and O.~Seto,
  Phys.\ Rev.\  D {\bf 82}, 023507 (2010)
  [\href{http://arxiv.org/abs/1002.2525}{arXiv:1002.2525} [hep-ph]];\\
  S.~Kanemura, O.~Seto and T.~Shimomura,
  Phys.\ Rev.\  D {\bf 84}, 016004 (2011)
  [\href{http://arxiv.org/abs/1101.5713}{arXiv:1101.5713} [hep-ph]].


\bibitem{micromegas}
  G.~Belanger, F.~Boudjema, A.~Pukhov and A.~Semenov,
  Comput.\ Phys.\ Commun.\  {\bf 176}, 367 (2007)
  [\href{http://arxiv.org/abs/hep-ph/0607059}{arXiv:hep-ph/0607059}];\\
  G.~Belanger, F.~Boudjema, A.~Pukhov and A.~Semenov,
  Comput.\ Phys.\ Commun.\  {\bf 180}, 747 (2009)
  [\href{http://arxiv.org/abs/0803.2360}{arXiv:0803.2360} [hep-ph]];\\
  G.~Belanger, F.~Boudjema, A.~Pukhov and A.~Semenov,
  \href{http://arxiv.org/abs/1005.4133}{arXiv:1005.4133} [hep-ph].


\bibitem{moreB-L}
  S.~Kanemura, T.~Nabeshima and H.~Sugiyama,
  Phys.\ Rev.\ D {\bf 85}, 033004 (2012)
  [\href{http://arxiv.org/abs/1111.0599}{arXiv:1111.0599} [hep-ph]];\\
  N.~Okada and Y.~Orikasa,
  \href{http://arxiv.org/abs/1202.1405}{arXiv:1202.1405} [hep-ph].


\bibitem{operatoranalysis}
  M.~Freytsis and Z.~Ligeti,
  Phys.\ Rev.\  D {\bf 83}, 115009 (2011)
  [\href{http://arxiv.org/abs/1012.5317}{arXiv:1012.5317} [hep-ph]].


\bibitem{xenon100} 
  E.~Aprile {\it et al.}  [XENON100 Collaboration],
  Phys.\ Rev.\ Lett.\  {\bf 107}, 131302 (2011)
  [\href{http://arxiv.org/abs/1104.2549}{arXiv:1104.2549} [astro-ph.CO]].


\bibitem{Geng:1988pr} 
  C.~Q.~Geng and R.~E.~Marshak,
  Phys.\ Rev.\ D {\bf 39}, 693 (1989).


\end{thebibliography}
\end{document}